\documentclass[runningheads]{svmult}
\usepackage{color}
\usepackage{makeidx}   
\usepackage{graphicx,amsfonts}  
\usepackage{subeqnar}  
\usepackage{multicol}  
\usepackage{physprbb}  
\makeindex             

\begin{document}
\title*{Nonextensivity: from low-dimensional maps to Hamiltonian systems}
\toctitle{Nonextensivity: from low-dimensional maps
\protect\newline to Hamiltonian systems }
%
%
%
\author{Constantino Tsallis\inst{1}
\and Andrea Rapisarda\inst{2}
\and Vito Latora\inst{2}
\and Fulvio  Baldovin\inst{1}
}

\authorrunning{Constantino Tsallis  et al.}

\institute{Centro Brasileiro de Pesquisas F\'{\i}sicas,\\
Rua Xavier Sigaud 150, 22290-180 Rio de Janeiro, RJ, Brazil
\and Dipartimento di Fisica e Astronomia
     and Infn Universit\'a di Catania,\\
     Corso Italia 57, 95129 Catania, Italy}

\maketitle             

\begin{abstract}
We present a brief pedagogical guided tour
of the most recent
applications of nextensive statistical mechanics to well defined nonlinear
 dynamical systems, ranging from one-dimensional
dissipative maps to  many-body Hamiltonian systems.
\end{abstract}

\section{Introduction}

Thermodynamics and Boltzmann-Gibbs (BG) statistical mechanics constitute
 central formalisms of  contemporary Physics. It is therefore of extreme
 importance to clearly establish under what exact conditions they are
 expected to apply, especially in what concerns the important and
 ubiquitous stationary state referred to as {\it thermal equilibrium}. This
 wisdom is more subtle than it looks at first sight. For instance,
 surprisingly enough, it is not yet exactly known the necessary and
 sufficient conditions for the celebrated BG distribution to correctly
 describe the thermostatistical state of a system (see, for instance,
 \cite{takensetal}). In order to settle
 such conditions, it is necessary to turn
 onto the microscopic dynamics of the system \cite{einstein}, would that be
 classical, quantum, 
relativistic, or any other one that  appropriately describes 
 the system under study.

Some of the central ingredients of 
BG statistical mechanics \index{statistical mechanics}
 are the well
 known logarithmic entropy ($S_{BG}=-k \sum_{i=  1}^W p_i \ln p_i$ in
 general, and $S_{BG}(\{p_i=  1/W\}) =   k \ln W$ for equiprobability, where
 for simplicity we are referring here to the case of discrete distributions
 of probabilities), and its exponential weight $p_i \propto e^{-\beta E_i}$
 ($\beta \equiv 1/kT$, $\{E_i\}$ being the energy spectrum) at equilibrium.
 This entropy has the remarkable property of extensivity \index{extensivity} 
 for systems which
 are independent. More precisely, if we have a system composed by subsystems
 $A$ and $B$ such that $p_{ij}^{A+B}=  p_i^A p_j^B$ ($\forall (i,j)$), then
 $S_{BG}(A+B) =   S_{BG}(A) + S_{BG}(B)$. This property is at the basis of
 the standard understanding of thermodynamics. Such a property is not
 necessarily universal  because it cannot be justified on general arguments
 (see, for instance, \cite{land}). Strong evidence exists
 nowadays that it should not be generically assumed, contrarily to a widely 
 spread, textbook belief among not few physicists. Very specifically, the 
 {\it Boltzmann principle} $S_{BG}(\{p_i=  1/W\}) =   k \ln W$ should not be 
 used for any given specific system unless properly justified. It was 
 Einstein \cite{einstein} the first who warned about this. However, many 
 physicists (see, for instance, \cite{gross}) have not yet done their way 
 through this important concept.  To be more precise, let us focus on 
 $N$-body classical Hamiltonian $d$-dimensional systems including two-body 
 interactions which do not present any mathematical complexity (such as an 
 infinitely attractive potential) at short distances. It is acquired by now 
 that the usual BG thermal equilibrium correctly describes the $t \to 
 \infty$ stationary state for all systems whose interactions are {\it 
 short-ranged} (e.g., if a two-body attractive potential decreases like 
 $1/r^\alpha$ with $\alpha/d>1$). In particular, in this case, it is 
 irrelevant to consider first the $t\to\infty$ limit and then the 
 $N\to\infty$ limit, or the other way around. The situation is much more 
 subtle if we have {\it long-ranged} interactions (e.g., $0 \le \alpha/d \le 
 1$). In this case, the $\lim_{N\to\infty} \lim_{t \to\infty}$ still leads 
 to the BG equilibrium, whereas the    $\lim_{t\to\infty} \lim_{N 
 \to\infty}$ generically does {\it not}. It happens, typically, that  more 
 than one basin of attraction exists in the space of the initial conditions. 
 For some initial conditions the system directly goes to the BG statioanary 
 state; for others it first goes to a non-BG state, remains there for a time 
 which diverges with $N$, and only eventually goes to the BG state.  In 
 other words, depending on the size of the system and on its specific 
 physical time scale, it might happen that the system remains in a 
 quasi-stationary state different from the BG one for times longer than the 
 age of the universe. In such cases, the only physically interesting 
 situation is the non-BG one, which, perhaps for a vast class of systems, 
 might be the one emerging within the nonextensive statistical mechanics we 
 shall describe in the next Section. The present paper is a tutorial review 
 of these concepts as they naturally emerge for three simple but important 
 classes of systems, namely, (i) one-dimensional dissipative maps (e.g., the 
 logistic map\cite{schuster,barangertsallis}), 
(ii) two-dimensional conservative maps (e.g., the standard 
 map\cite{stand}), and (iii) classical many-body 
long-range-interacting Hamiltonians 
 (e.g., the HMF model \cite{ant1,lat1,lat2,lat3,hmfequi} 
and one of its generalizations, the 
 so called $\alpha-XY$ model \cite{celia,gia}).
We shall not address here the many applications of nonextensive statistical 
 mechanics available in the literature. The interested reader may refer to 
 \cite{tsallis2} for various reviews. Just as an illustration, let us 
 mention some of those applications: turbulence 
 \cite{bect,arimitsu}, electron-positron annihilation 
 \cite{bediaga}, diffusion of {\it Hydra viridissima} \cite{arpita}, 
 diffusion of quarks in gluon plasma \cite{rafelski}, Levy and correlated 
 anomalous diffusions \cite{levyetal}, linguistics \cite{montemurro}, 
 economics \cite{borland,celiaconstantino}, fluxes of cosmic rays 
 \cite{cosmic}, solar neutrinos \cite{quarati}, high energy particle 
 collisions \cite{ion}, self-organized criticality \cite{SOCtamarit}, among 
 others. In one way or another, these phenomena seem to share long-range 
 correlations in space/time, either long-range microscopic interactions, or 
 long-range microscopic memory (nonmarkovian processes), or (multi)fractal 
 boundary conditions, or, generically speaking, some mechanism which creates 
 a scale-invariant hierarchical structure of some sort.   Such possibility 
 appears to emerge in classical many-body Hamiltonians and other nonlinear 
 dynamical systems, everytime the Lyapunov spectrum approaches zero, hence 
 chaos becomes impossible in the sense that the sensitivity to the
 initial conditions diverges less than exponentially in time, though a weak
mixing can remain.

\section{Mathematical formalism}

The first and main step in order to go from BG to nonextensive 
\index{non extensive thermodynamics}
 thermostatistics is to propose the use of an entropic form which 
 generalizes that of BG, as follows \cite{tsallis1}:
\begin{equation}
S_q\equiv k\frac{1-\sum_{i=  1}^W p_i^{\;q}}{q-1}\;\;\;\;(\sum_{i=  1}^W p_i 
 =  1;\;q \in \mathbb{R})\; .
\label{stsallis}
\end{equation}
(for simplicity, and without loss of generality, we shall adopt $k=1$ from 
 now on). This nonnegative form can be conveniently rewritten as
\begin{equation}
S_q =   \langle \ln_q \frac{1}{p_i} \rangle \;,
\end{equation}
where $\langle...\rangle \equiv \sum_{i=  1}^W (...) p_i$, and the {\it 
 $q$-logarithm} function is defined as
\begin{equation}
\ln_q \equiv \frac{x^{1-q}-1}{1-q} \;\;(\ln_1 x=  \ln x;\; x \ge 0) \;.
\end{equation}
The inverse function is the {\it $q$-exponential} one, given by
\begin{equation}
\label{qexp}
e_q^x \equiv [1+(1-q)\; x ]^{\frac{1}{1-q}}\;\;(e_1^x =  e^x) \;,
\end{equation}
solution of $dy/dx =   y^q$ with $y(0)=  1$.
$S_q$ satisfies several remarkable properties, such as {\it concavity} for 
 all $\{p_i\}$ ($\forall q>0$), {\it stability} with regard to $\{p_i\}$ 
 ($\forall q>0$) \cite{abestability}, and, for $A$ and $B$ independent 
 subsystems, the following {\it pseudo-extensivity}:
\begin{equation}
S_q(A+B) =   S_q(A)+S_q(B)+(1-q)S_q(A)S_q(B) \;,
\end{equation}
from which the denomination {\it nonextensive} comes. For completeness, it 
 is convenient to mention at this point that similar entropies are the Renyi 
 one, defined as $S_q^R =   [\ln \sum_{i=  1}^W p_i^q]/[1-q]$, and the 
 normalized one $S_q^N=  S_q / [\sum_{i=  1}^W p_i^q]$ (introduced 
 independently by Abe and Rajagopal \cite{normalized1}
and by Landsberg and Vedral \cite{normalized2}. Renyi entropy is very useful 
 in the context of multifractal analysis (of nonlinear chaotic systems, for 
 example). However, $S_q^R$ and $S_q^N$ are {\it not} concave for all 
 positive values of $q$, {\it nor} are they stable \cite{abestability}. 
 These facts constitute severe drawbacks for basing a thermodynamical 
 formalism on them. It is however convenient to have in mind that, since 
 they are related to each other through monotonic functions 
$S_q^R=  \ln [1+(1-q)S_q]/[1-q]$ and $S_q^N=  S_q/[1+(1-q)S_q]$, 
they become extremized 
 by the {\it same} probability distributions (assuming the supplementary 
 constraints for optimization to be the same).

\section{Applications to low-dimensional maps}

Low-dimensional systems are
a field in which the nonextensive statistical mechanics has found one of
its first and most proficuous applications. 
What makes low-dimensional systems extremely attractive is
the simplicity of their dynamics and the possibility to have
direct access to the phase space, sometimes even to plot
a visual image of the entire phase space. In fact,
the iterative rules of low-dimensional maps 
can be easily implemented on a computer and have provided  
useful methods to perform ``numerical experiments'' to check for
the validity of the nonextensive formalism.
As mentioned  in the previous section, the BG statistics is expected
to fail when the sensitivity to initial conditions is not exponential.
Therefore, different authors have looked and found applications
of the nonextensive formalism to
cases in which the maps have a nonstandard
sensitivity to initial conditions, 
as indicated by the vanishing of the Lyapunov exponents.
Different techniques have been produced in order to
calculate the `entropic parameter' $q$, and it turns out
that this parameter characterizes `classes of universality'
of systems, at least for some of their statistical behaviors.
For pedagogical reasons here we will discuss only some of
the results obtained for the logistic map,
as representative of a dissipative system, and for the
standard map, as an example of a conservative one.

\subsection{Dissipative Maps: the logistic map}

The logistic map is one of the simplest \index{logistic  map} 
one-dimensional dissipative
system one can imagine. It has been intensively
studied and it has led to important developments in chaos 
theory \cite{schuster}.
Since it has  all fundamental characteristics of
non-conservative systems, the logistic map is often taken 
as a paradigmatic example;
in other words it can be considered as the Ising model of 
non-linear dynamical systems.
The logistic map can be described by the following iterative
rule\footnote{An equivalent, perhaps more popular version
of the logistic map is
$x_{t+1} =   rx_t(1-x_t)$,
 with $0\le x_t \le1$ and $1\le r \le4$.
However, we notice that form (\ref{log}) is easily generalized to
other universality classes of order $z\geq 1$:
$x_{t+1} =   1-a|x_t|^z$.
In this way, a $z$-logistic map is in fact representative
of the universality
class of unimodal maps of order $z$; i.e., of one-dimensional maps with
a single maximum of order $z$.
}
\begin{equation}
x_{t+1} =  f_a(x_t) =   1-a x_t^2\;\;\; -1 \le x_t \le1; \;0\le a \le2;\;\;
\label{log}
\end{equation}
and it shows different regimes according to
the value of the control parameter $a$ \cite{schuster,beck}.
\begin{figure}[th]    
\begin{center}
\includegraphics[width=0.65\textwidth]{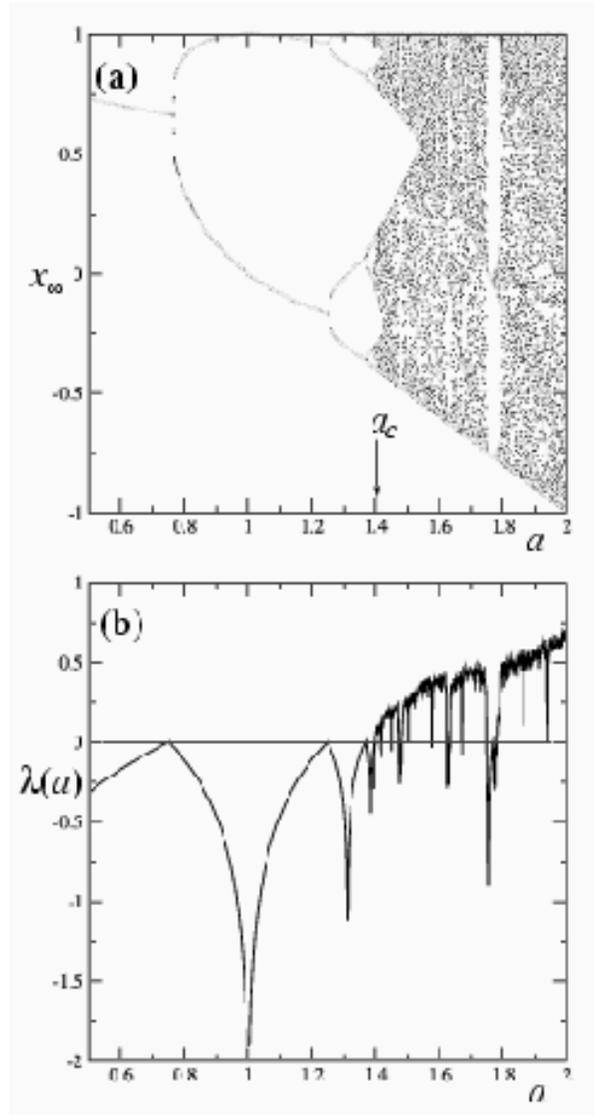}
\end{center}
\caption[]{
(a) Attractor of the logistic map as a function of $a$.
The edge of chaos \index{edge of chaos} 
is at the critical value $a_c=  1.401155198...$
(b) Lyapunov exponent $\lambda$ as a function of $a$.}
\label{log_fig}
\end{figure}
For small values of $a$ the map is regular since it has non-positive
Lyapunov exponents (see Fig.\ref{log_fig}).
In particular, for $a<\frac{3}{4}$ the iterates of the logistic map
converge to a single fixed point $x^* =   f_a(x^*)$,
whose value depends on $a$.
At $a=  \frac{3}{4}$ we have $\lambda=  0$ and
we observe the first pitchfork bifurcation:
for a value slightly larger than $\frac{3}{4}$ the iterates
converge to a cycle-2.
The number of  points of the attractor $f_a(x)$ keeps doubling at distinct,
increasing values of the parameter $a$, until
a critical value $a_c=  1.401155198...$ is reached, where
the attractor becomes infinite. 
This is known in the literature as the period-doubling route
to chaos. In fact, beyond $a_c$
the behavior of the logistic map is chaotic (positive
Lyapunov exponent $\lambda$) for most of the values of $a$.
At the chaos threshold $a=  a_c$, $\lambda$ is once again zero,
and this point is the well known {\it edge of chaos} 
which has been the subject of intensive debates about the
 definition of what a complex systems is
\cite{SOC,barcomplex}. It is exactly at the edge of chaos
and at all other points where $\lambda$ is zero (as at
period doublings,
but also for instance at inverse tangent bifurcations
\cite{schuster}),
that a generalization of the standard BG entropy
will be fruitful, as we show below.

\subsubsection{Sensitivity to initial conditions}
\begin{figure}[ht]    %
\begin{center}
\includegraphics[width=.65\textwidth]{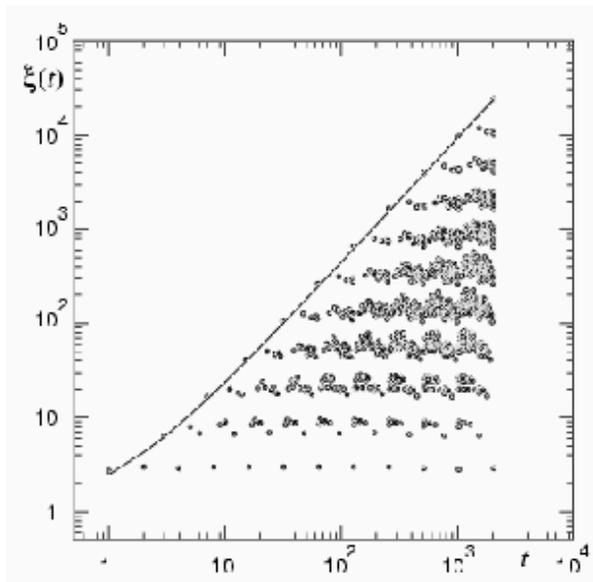}
\end{center}
\caption[]{Sensitivity to initial conditions for the logistic
map at $a=  a_c$. The dots represent $\xi(t)$ for
two initial data started at
$x_0=  0$ and $x_0^\prime\sim 10^{-8}$.
The solid line
is the function in Eq. (\protect\ref{q-lya}), with
$q=  0.2445...$ and $\lambda_q= \ln \alpha / \ln 2=  1.3236...$. }
\label{monster}
\end{figure}

Let us start by explaining better the meaning of the crucial
concept of sensitivity to initial conditions. 
For all the cases in which the Lyapunov exponent $\lambda$ is
positive (negative) we expect on the average an exponential increase
(decrease) of any small initial distance
$\xi(t) \equiv \frac{|\mathbf x_t-\mathbf x_{t}^{\prime}|}
{|\mathbf x_0-\mathbf x_{0}^{\prime}|}$
to hold. We typically have
\begin{equation}
\xi(t)=  \exp{(\lambda t)},
\label{lya}
\end{equation}
where $\mathbf x_t$ and $\mathbf x_{t}^{\prime}$ are the positions 
at time $t$ of two initially close trajectories
that may be considered, for the sake of generality, in a
$d$-dimensional phase space.
In the case of the logistic map $d=1$.
This case is
referred in the literature as
{\it strongly sensitive} ({\it insensitive}) 
to initial conditions.
What is not usually reported in textbooks is what happens
when $\lambda$, as defined by equation (\ref{lya}),
is equal to zero.
In Ref. \cite{tsallisplastino} was proposed that
in such situations
the system still exhibits sensitivity to initial conditions,
though in a form that is the $q$-generalization of Eq. (\ref{lya}):
\begin{equation}
\xi(t)=  \exp_q(\lambda_q t)\equiv
[1+(1-q)\lambda_q t]^{\frac{1}{1-q}}\;\;\;(q \in \mathbb{R}).
\label{q-lya}
\end{equation}
This equation recovers Eq. (\ref{lya}) for $q=  1$, and displays
a quite rich spectrum of possiblities according to the sign
of $\lambda_q$ and $q$.
Let us concentrate on the case $\lambda_q>0$,  $q<1$.
In this case the system presents a {\it weak sensitivity}
to initial conditions, in the sense that two
initially close trajectories diverge in time as a power-law
instead than exponentially.
A weakly sensitive system is well described once $q$ and
$\lambda_q$ are given.
In \cite{tsallisplastino} $q$ was calculated numerically
for the logistic map at the edge of chaos,
observing that the upper bound of $\xi(t)$ lies on a
line that is precisely a power-law (see Fig. \ref{monster});
furthermore, it was conjectured that a value  $q=0.2445....$ 
could be deduced
from the Feigenbaum's constant $\alpha$.
These results were recently confirmed in \cite{robledo_01},
where it was proved that, at the edge of chaos \index{edge of chaos}
the upper bound of $\xi(t)$ has exactly the form of Eq. (\ref{q-lya}),
with $q=0.2445...$ and $\lambda_q= \ln \alpha / \ln 2 =1.3236...$.
In fact, as a consequence of the Feigenbaum's
renormalization group recursion
relation, two close initial conditions starting in the neighborhood of
the origin $x=0$, produce, each $t=2^n-1$ ($n=  0,1,2,...$),
a dominant power-law separation of their
iterates that has the form (\ref{q-lya}). After these particular
steps, the iterates reproduce self-similar sequences,
that are also power-laws. The effect of starting the initial
points in other regions of the phase space instead that in the
vicinity of $x=0$ is just a shifting in time of Fig. \ref{monster}.

The renormalization group approach has proven powerful also at other
points where the Lyapunov exponent is zero. In \cite{robledo_02}
the same $q$-exponential form (\ref{q-lya}) has been exactly obtained
for the sensitivity to initial conditions at pitchfork and inverse
tangent bifurcations, now with other values for $q$ and $\lambda_q$.
It is important to notice that the previous scheme of application
of Eq. (\ref{q-lya}) generalizes to the other universality
classes of unimodal maps of order\footnote{See previous footnote.} $z>1$.

\subsubsection{Entropy production}
\begin{figure}[ht]    
\begin{center}
\includegraphics[width=0.75\textwidth]{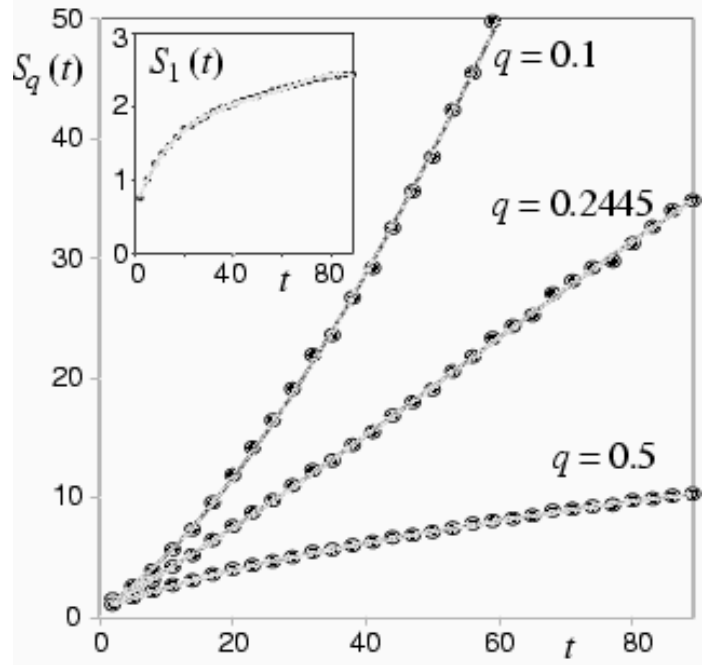}
\end{center}
\caption[]{Time evolution of $S_q$ for the logistic map with $a=  a_c$. 
We consider four different values of $q$.
The case $q=  1$ is in the inset.  $W=  N=  2.5\;10^6$
and results are averages over 15115 runs.}
\label{s_log2}
\end{figure}

The analysis of the entropy production provides
an alternative approach to the same problem and
a second way
to characterize the $q$-index at the edge of chaos.
Instead of
considering two trajectories initially close, we take a
distribution of initial conditions localized
in a tiny region of the phase space, we let this distribution evolve 
in time according to Eq. (\ref{log}), and we study
the increase of the entropy \cite{baranger1,barangertsallis}.
In the case of the logistic map,
we can partition the phase space interval
$-1 \le x \le 1$ into $W$ equal cells, and consider
an initial distribution of $N$ points placed at random
inside one of the cells; the initial cell is
also chosen randomly in the partition.
The normalized number of points that occupy a cell defines a probability
distribution: $p_i\equiv n_i/N$ ($\sum_{i=  1}^W p_i=  1$),
and as the system evolves, from the probabilities $p_i(t)$ we can
calculate the entropy production $S_q(t)$, via Eq. (\ref{stsallis}).
In the chaotic regime the BG entropy ($q=1$)
exhibits, before a saturation due to the finiteness of $W$,
a linear increase in time, and in this stage the entropy production rate
(that is expected to be equal to the Kolmogorov-Sinai entropy \cite{kolmo})
coincides with the positive Lyapunov exponent \cite{pesin}.
\begin{figure}[ht]    
\begin{center}
\includegraphics[width=0.65\textwidth]{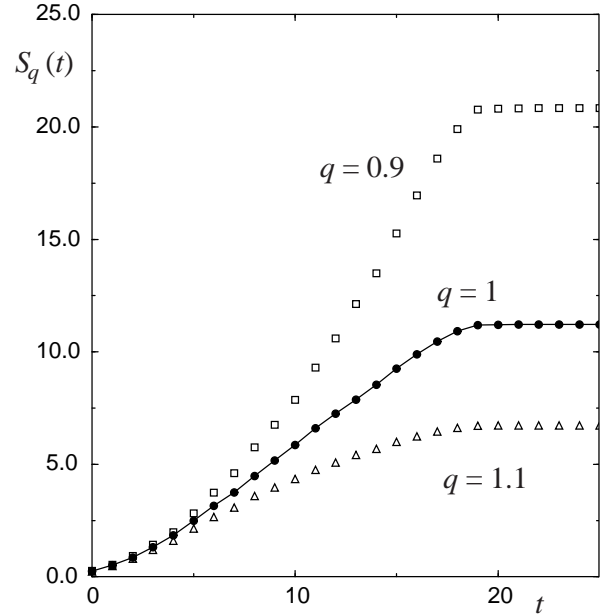}
\end{center}
\caption[]{Time evolution of $S_q$ for the logistic map with $a=2$. 
We consider three different values of $q$.  
The results are averages over 500 runs with $N= 10^6$ and 
$W= 10^5$.}
\label{s_log1}
\end{figure}
This is shown in Fig. \ref{s_log1} for the case $a=  2$, where
$S_q(t)$ is reported for three different values of $q$.
As $t$ evolves, $S_q(t)$ increases (in all cases bounded by
$\frac{W^{1-q}-1}{1-q}$, or $\ln W$ when $q=  1$), but only the curve
for $q=  1$ shows a clear linear behavior with a
slope equal to the Lyapunov exponent $\lambda=  \ln 2$ \cite{baranger1}. 
For $q<1$ the curve is convex, while for $q>1$ the curve is concave.
The slope in the linear stage does not depend
on the dimension of the cells of the partition and on the
size of the initial distribution.
Therefore, in the chaotic regime the standard B-G entropy is
the only $q$-entropy that displays a finite entropy production
rate $\lim_{t\to\infty\;W\to\infty}S_q(t)/t$.
We wish now to repeat the same analysis at the edge of
chaos, where $\lambda$ is zero. 
Due to the  power-law
sensitivity to initial conditions and to the fractality
\index{fractals}
of phase space (the attractor now is a multifractal)
we expect that anomalous behavior may be observed.
In Fig. \ref{s_log2} we consider $a=  a_c$ and we plot $S_q(t)$ for
four different values of $q$;
the curves are obtained with a large $W$ and with
an average over many different runs since
the complexity of the phase space originates many fluctuations.
Consistently with the value of $q$ extracted for the sensibility to initial
conditions, the result is that
the growth of $S_q(t)$ is found to be linear
when $q=0.2445...$, while for $q<q_c$ ($q>q_c$) the curve is convex
(concave). 
This behavior is similar to the one of Fig. \ref{s_log1},
with a major difference: the linear growth is not at $q=  1$,
but at $q=  0.2445...$ \cite{barangertsallis}.

\subsubsection{Multifractal analysis}
There is also a third, geometrical method that allows us to calculate
the $q$ of the logistic map. This method is based only on the 
description of the multifractal attractor existing at $a=a_c$,
and gives exactly the same value we have already obtained.  
The multifractal attractor can be characterized by
using the multifractal function $f(\alpha)$ \index{fractals}
\cite{beck}. This function is defined in the interval
$[\alpha_{min},\alpha_{max}]$, and its maximum equals the fractal
or Hausdorff dimension $d_f$.
The value of $q$ can be calculated from \cite{tsallisplastino}:  
\begin{equation}
\frac{1}{1-q}=   \frac{1}{\alpha_{min}}-\frac{1}{\alpha_{max}}~,
\end{equation}
where $\alpha_{min}$ ($\alpha_{max}$) 
are the two values for which $f(\alpha)$ is defined.
In particular for the logistic map at the edge of chaos
$\alpha_{min}=  0.380...$, $\alpha_{max}=  0.755...$ and
we have again $q=0.2445....$

\subsection{Conservative Maps: the standard map}

\begin{figure}[ht]    
\begin{center}
\includegraphics[width=0.75\textwidth]{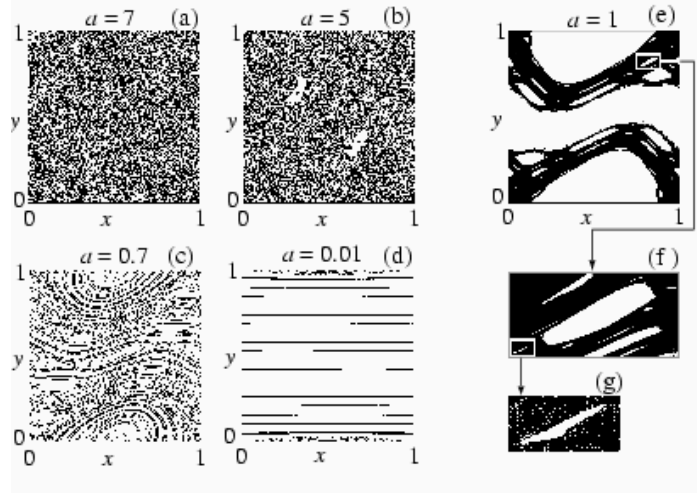}
\end{center}
\caption[]{(a)--(d) Phase portrait of the standard map for
typical values of $a$. $N=  20\times20$ orbits (black dots)
were started with a uniform  distribution in the unit square
and traced for $0\leq t\leq 200$.
(e)--(g) Islands-around-islands. (e): $N=  100\times100$ initial
data were started inside a small square of side $\sim 10^{-2}$
and traced for $0\leq t\leq 5000$. (f) is a magnification
of the island inside the rectangle in (e).
(g) is a magnification
of the island inside the rectangle in (f).
}
\label{islands}
\end{figure}

Conservative systems present quite different statistical
properties than dissipative ones.
One of these differences is that \index{standard map}
the transition from chaoticity to regularity
happens in the phase space without an `edge of chaos',
but in a rather complicated way as described by
the Kolmogorov-Arnold-Moser (KAM) theorem 
\cite{ott_01}, see Fig. \ref{islands}e-g).
Conservative maps may be derived, for example,
from Hamiltonian
systems. In this case the original system is characterized
by an even number of dimensions $d=2n$, where $n$ is the
number of degrees of freedom.
If the Hamiltonian $H$ is time-independent the dimension can be reduced by one.
Moreover, as in statistical mechanics we are interested
mostly in {\em recurrent trajectories}
(i.e., in those trajectories that come back again and
again, indefinitely, to any part of the phase space they
have once visited), we can take a {\em Poincar\'e section}
of the phase space cutting transversally the constant-energy
hypersurface and considering the successive intersections
of each orbit with this transversal surface.
In this way we reduce the dimension of the phase space to
$d_M=  2n-2$ and we obtain a great  numerical simplification,
as we can now analyze a discrete-time iteration map
instead of a continuous-time system of differential equations.
The map thus obtained may be shown to be {\em simplectic}
(see, e.g. \cite{ott_01}); this implies that the $d_M $ Lyapunov exponents
are coupled in pairs, where each member of the pair has  the
opposite of the other.

The minimum dimension that allows for the application of this
scheme is $d_M=  2$ ($n=  4$), and a paradigmatic
example of this kind, that plays for conservative
systems a similar role the logistic map plays for
dissipative ones, is the standard map, that may be
characterized by the equations\footnote{The map is also
known as the kicked rotator map.}:
\begin{eqnarray}
x_{t+1}&=  &y_t+\frac{a}{2\pi}\sin(2\pi x_t)+x_t~~~\textup{(mod
1)},\nonumber\\ \\
y_{t+1}&=  &y_t+\frac{a}{2\pi}\sin(2\pi x_t)~~~~~~~~~\textup{(mod
 1)}\nonumber,
\label{standard}
\end{eqnarray}
where $a\in\mathbb{R}$ ($t=  0,1,...$).
The system is integrable
when $a=  0$, while,
roughly speaking, chaoticity increases when $|a|$
increases (see Fig. \ref{islands}a-d).

As for the logistic map, we expect that anomalous
effects may be produced as the complexity of the
phase space increases (i.e., when we pass from the chaotic
to the regular regime reducing $|a|$), and as this happens,
we would like to see how the nonextensive  formalism generalizes
the BG one, at least for some of the statistical
variables we have previously described.

\subsubsection{Sensitivity to initial conditions}
\begin{figure}[ht]    %
\begin{center}
\includegraphics[width=0.65\textwidth]{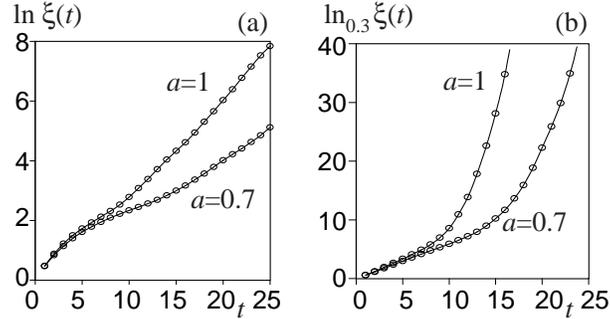}
\end{center}
\caption[]{
Average sensitivity to initial conditions
of the standard map for small values of $a$. 
$|\mathbf x_0-\mathbf x_{0}^{\prime}|\sim 10^{-9}$
and results are averages over 5000 runs.
An initial separation of order
In (a) $\ln(\xi(t))$
displays a linear increase only after a certain
time. Correspondingly, in (b) $\ln_{0.3}(\xi(t))$
exhibits a linear increase for the initial steps.
}
\label{xi_std}
\end{figure}

In the case of the standard map,
the sensitivity to initial condition depends dramatically
on the position in the phase space. Chaotic regions produces
exponential separation of initially close trajectories, while
regular regions exhibit linear separation of initially close
orbits. At the border between these regions,
appear fractal-like
structures of islands-around-islands
(see, e.g., \cite{zaslavsky_01}, see Fig. \ref{islands}e-g).
We notice however that in statistical mechanics we are mainly interested
in extracting global averages behaviors, so that we can try to calculate
an average $\xi(t)$ by sampling uniformily the whole phase space.
When this experiment is performed for rather small values of $|a|$,
as we  show in Fig. \ref{xi_std}, a {\em crossover}
in time between two different regimes occur \cite{tsallis_baldovin_01}. 
Fig. \ref{xi_std}a displays in fact a linear 
increase of $\ln(\xi(t))$ only after a certain 
iteration step. Correspondingly, Fig. \ref{xi_std}b
exhibits a linear increase of $\ln_q(\xi(t))$ precisely for
those steps where $\ln(\xi(t))$ is concave
(numerically it was found $q\simeq 0.3$).
Moreover, the  crossover time increases when $|a|$ decreases.
This means that two different behaviors
dominates the average sensitivity to initial conditions
for different time. Initially,
because of the predominance of fractal-like structures in
the phase space, the
sensitivity to initial conditions is power-law; then, after
a certain time, the sensitivity to initial conditions
becomes exponential due to the rapidity of the exponential
growth if compared with the power-law.
If we wish to model such a behavior with a single function, we
can first observe that exponential and power-law sensitivity
to initial conditions may be viewed as solution of the differential
 equations
$\dot\xi=  \lambda\xi$ and
$\dot\xi=  \lambda_q\xi^q$ respectively.
A differential equation describing a crossover in time is then:
$\dot\xi=  \lambda\xi+(\lambda_q-\lambda)\xi^q$
($q<1,\;\;\lambda,\lambda_q>0$).
The solution \cite{tsallis_bemski_01}
\begin{equation}
\xi(t)=  \left[1-\frac{\lambda_{q}}{\lambda}+\frac{\lambda_{q}}{\lambda}
e^{(1-q)\lambda t}\right]^{\frac{1}{1-q}},
\label{q-cross}
\end{equation}
presents, in the case $0<\lambda\ll\lambda_{q}$, three asymptotic
behaviors, namely (i) linear:
$\xi\sim 1+\lambda_{q}t$, for $0\leq t\ll
t_{cross1}\equiv\frac{1}{(1-q)\lambda_{q}}$;
(ii) power-law:
$\xi\sim\left[(1-q)\lambda_{q}\right]^{\frac{1}{1-q}}
t^{\frac{1}{1-q} } $, for $t_{cross1}\ll t\ll t_{cross2}\equiv
\frac{1}{(1-q)\lambda}$; (iii) exponential:
$\xi\sim (\frac{\lambda_q}{\lambda})^{\frac{1}{1-q}}e^{\lambda t}$,
for $t\gg t_{cross2}$.

\subsubsection{Entropy production}

\begin{figure}[ht]    %
\begin{center}
\includegraphics[width=0.65\textwidth]{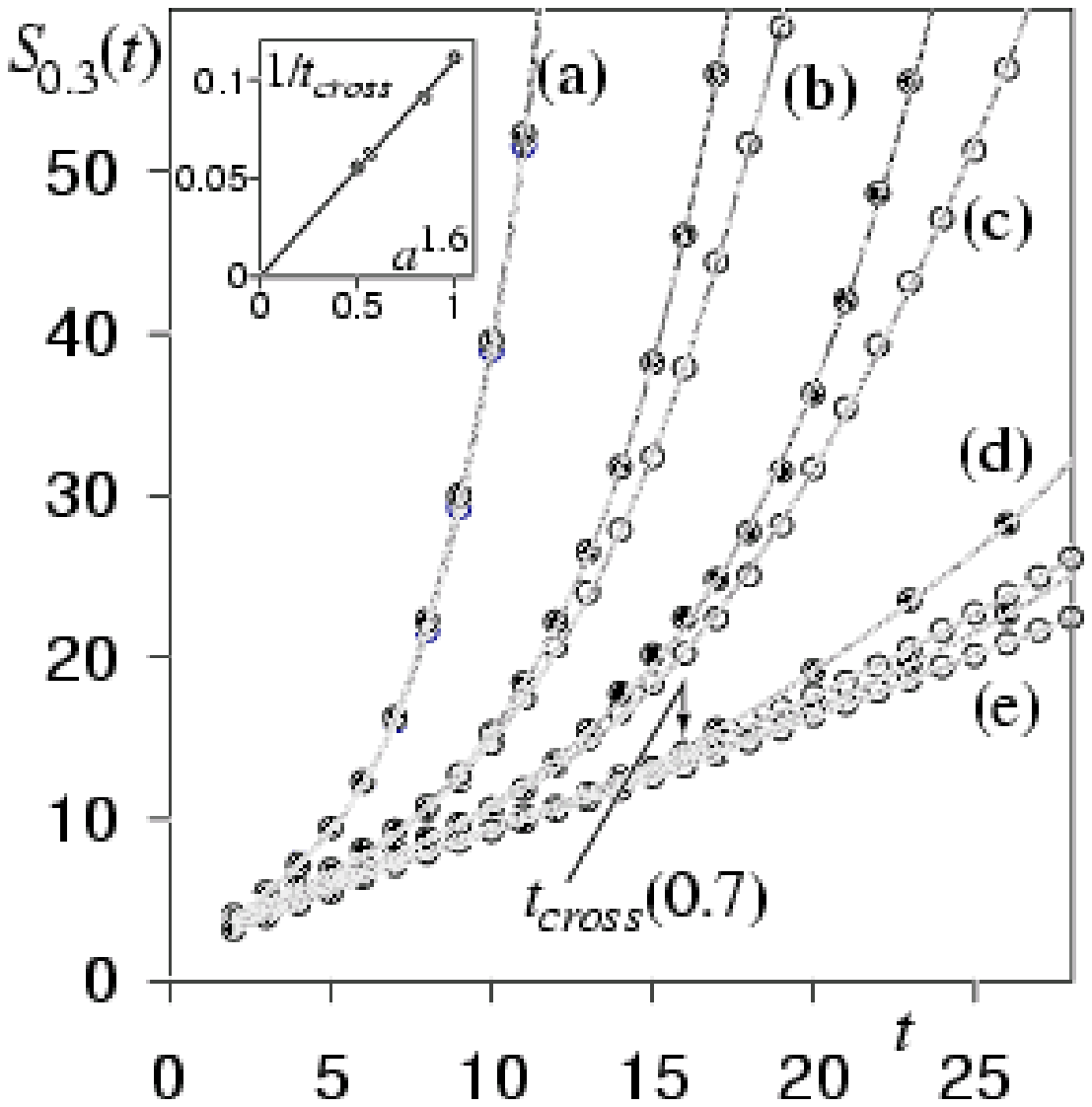}
\end{center}
\caption[]{For the standard map we show
$S_{0.3}(t)$ for: $a=  1.5$ (a); $a=  1.1$ (b);
$a=  0.9$ (c); $a=  0.7$ (d); $a=  0.65$ (e)
($5000$ to $7000$ runs were averaged). Full circles  correspond to
 $N=  W=  1000\times 1000$
for (a), (b), (c),
$N=  W=  5000\times 5000$ for (d), (e);
empty circles correspond to $N=  W=  448\times 448$ for (a), (b), (c),
$N=  W=  2236\times 2236$ for (d), (e).
Inset: $t_{cross}(a)$ defined as the intersection of the linear part (before
 it starts bending)
and a standard extrapolation of the bended part of the curves $S_{0.3}(t)$.

Notice that $t_{cross}(a)$
diverges for $a\to 0$.
The lines are guides to the eye.
Our results suggest
$\lim_{t\to\infty}\lim_{a\to
 0}\lim_{W\to\infty}\lim_{N\to\infty}\frac{S_{0.3}(t)}{t}\simeq 0.71$
for $q=  q^*\simeq 0.3$, whereas this limit vanishes (diverges) for $q>q^*$
 ($q<q^*$).
}
\label{s03_std}
\end{figure}

If the connection between sensitivity to initial condition
and entropy production \index{entropy production} 
that we have observed for the
logistic map is correct, we should observe a
crossover between different statistical regimes even when
we analyze the entropy production of the standard
map. This has been done in \cite{tsallis_baldovin_02},
with a numerical experiment analogous to the one previously
described for the logistic map and that is difficult to perform
because of the complexity of the phase space and the
related
effects of rapid saturation of the entropy for $W$ finite.
\begin{figure}[ht]    
\begin{center}
\includegraphics[width=0.65\textwidth]{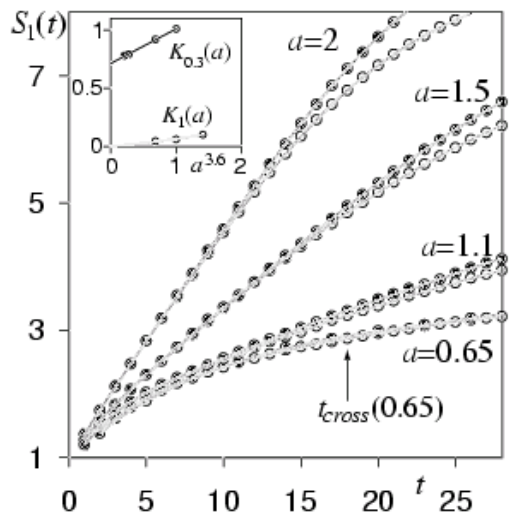}
\end{center}
\caption[]{For the standard map we show 
$S_1(t)$ for different  values of $a\leq 2$
($5000$ to $7000$ runs were averaged). Full circles  correspond to
 $N=  W=  1000\times 1000$
for $a=  2,1.5,1.1$
($N=  W=  5000\times 5000$ for $a=  0.65$);
empty circles correspond to $N=  W=  448\times 448$ for $a=  2,1.5,1.1$
($N=  W=  2236\times 2236$ for $a=  0.65$).
Inset: Slopes of $S_1(t)$ ($K_1(a)$) and of $S_{0.3}(t)$ =20
($K_{0.3}\simeq 0.71+0.30 \;a^{3.6}$)
in their linear regimes (see also Fig. \ref{s03_std}).
The lines are guides to the eye.
}
\label{s1_std}
\end{figure}
In Figs. \ref{s1_std} and \ref{s03_std} we display
how, for small values of $|a|$ and before saturation
begins, the entropy production of $S_1(t)$
becomes linear only after a stage where $S_q(t)$ is linear, with,
once again, a value $q\simeq 0.3$.
Notice that the crossover time tends to infinity
when $|a|$ goes to zero, so that in this limit the stage
characterized by the power-law sensitivity to initial conditions
and by this nonstandard entropy production lasts for infinite time.

\section{Applications to Many-particle systems}

\subsection{Hamiltonian nonextensive systems: the HMF model }

In the last years there has been an intense research activity in investigating 
microscopic dynamical features in connection with macroscopic features.
 In particular
several authors have focused their 
attention in characterizing from a dynamical point
of view macroscopic phase 
transitions \cite{bona,pettini,pos,kon,tor}. 
It has been found, for example, that in correspondence 
of a phase transition there \index{phase transition} 
is a peak in the chaoticity of the microscopic dynamics.
In general chaoticity is of 
fundamental importance for  any system to reach the standard equilibration.
Anomalies should be expected when 
this does not occur, 
since ergodicity breaking might 
happen and 
the system can remain trapped in some  region of phase space. 
In this respect a model whose 
features have been particularly interesting is the Hamiltonian
Mean Field model, \index{mean field models} from here on 
referred as HMF model, and its generalizations where the range of the 
interaction can be changed. 
The model 
was introduced   in ref \cite{ant1} and 
extensively studied in refs \cite{ant1,lat1,lat2,lat3,celia,gia}. 
In this volume one 
chapter is dedicated to its equilibrium 
properties both from a dynamical and from a thermodynamical 
point of view \cite{hmfequi}. 
In the present  section we consider only the 
anomalies found in the out-of-equilibrium
regime before equilibration \index{metastability} 
to the Boltzmann-Gibbs regime and 
how these behavior can be related to nonextensive statistics \cite{tsallis1}.

\begin{figure}[t]
\begin{center}
\includegraphics[width=.75\textwidth]{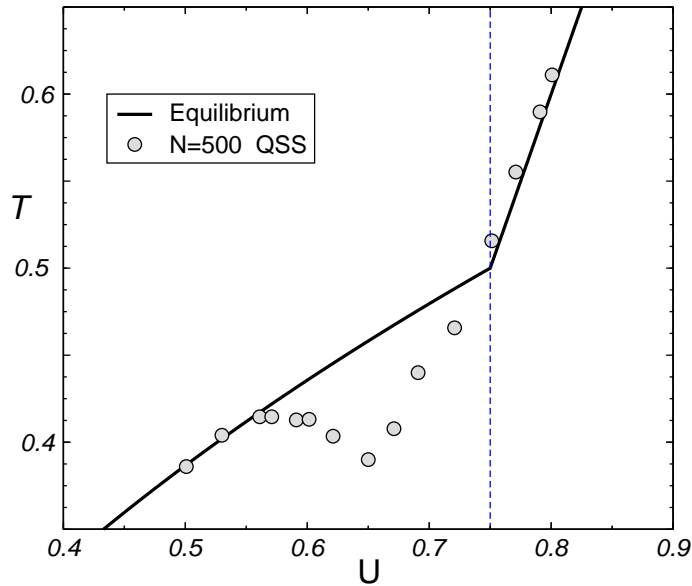}
\end{center}
\caption[]{We compare the equilibrium caloric curve of the HMF model
(full curve) with the 
numerical simulation for a system 
of size N=500. The numerical integration is followed up to 
 a short time, when the system is 
in the metastable quasi-stationary  state (QSS). }
\label{calor}
\end{figure}

\begin{figure}[ht]
\begin{center}
\includegraphics[width=.75\textwidth]{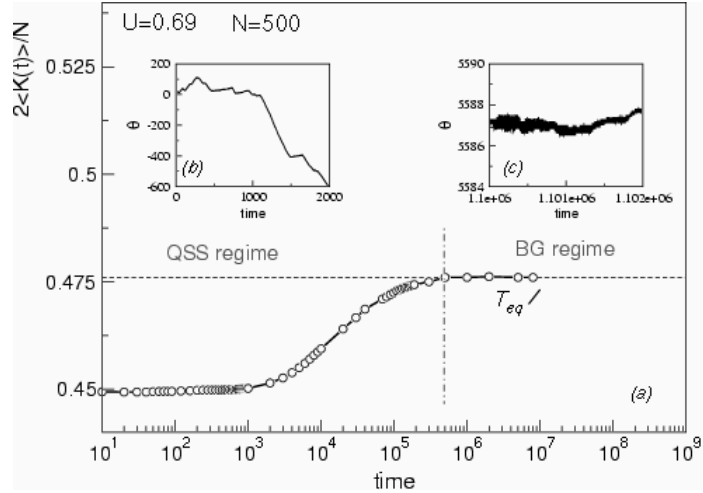}
\end{center}
\caption[]{For the HMF model at  U=0.69 and N=500, we plot the temperature, 
calculated by means of the average 
kinetic energy, as a function of time. 
Two plateaux emerge from the calculation. The 
first one is the metastable 
quasi-stationary state (QSS), 
while the second one corresponds to the Boltzmann-Gibbs
(BG) equilibrium regime.  In the inset (b) we show the angle vs
 time of a typical rotor in the 
metastable regime: the motion
shows  L\'evy walks and 
anomalous diffusion \protect\cite{lat2}. In the inset (c) we show the
single-particle motion in the 
BG equilibrium regime: the motion is a random walk.   }
\label{voliqss}
\end{figure}

\begin{figure}[ht]
\begin{center}
\includegraphics[width=.75\textwidth]{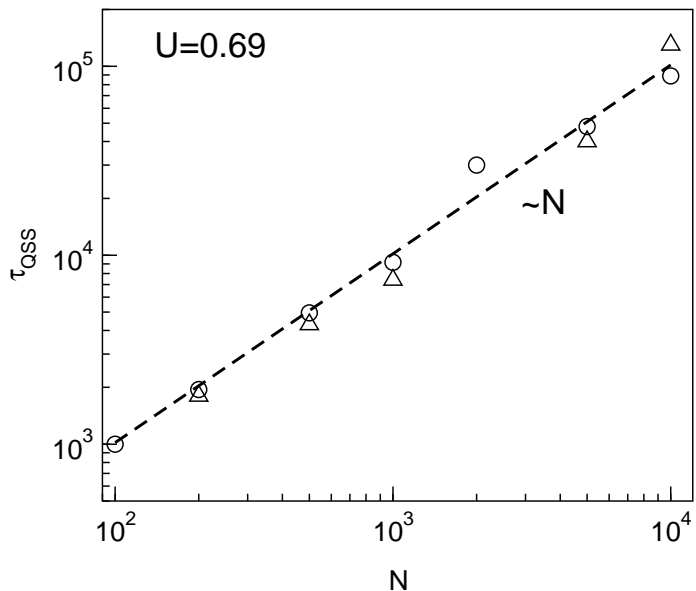}
\end{center}
\caption[]{For the HMF model we plot the lifetime of 
the metastable quasi-stationary state, $\tau_{QSS}$, 
for U=0.69, as a function of N.
Different out-of-equilibrium 
initial conditions are indicated by different symbols. The figure
shows that the duration of these states diverges with N.   }
\label{scaling}
\end{figure}

\begin{figure}[ht]
\begin{center}
\includegraphics[width=.75\textwidth]{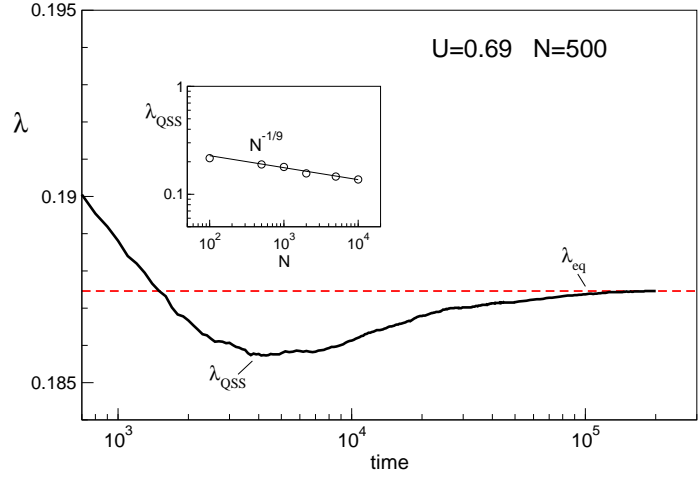}
\end{center}
\caption[]{We plot the largest Lyapunov exponent, $\lambda$,
 vs time for the HMF model at $U=0.69$ 
and $N=500$. Two plateaux are found also in this 
case. In the inset we show that $\lambda_{QSS}$ goes to zero 
as $N^{-1/9}$. This behavior can be 
explained with the anomalous fluctuations 
of the magnetization. See text for further details.  }
\label{lle}
\end{figure}

The HMF  model describes a system of 
N planar classical spins interacting through an 
infinite-range potential\cite{ant1}.  
The Hamiltonian can be written as:
\begin{equation}
        H=K+V= \sum_{i=1}^N  {{p_i}^2 \over 2} +
  {1\over{2N}} \sum_{i,j=1}^N  [1-cos(\theta_i -\theta_j)]~~,
\label{hmfpot}
\end{equation}
\noindent
 where $\theta_i$ is the $ith$ angle and $p_i$ the 
conjugate variable   representing   the  angular momentum
or the rotational velocity since unit mass is assumed. 
The interaction is the same as in the
ferromagnetic XY model \cite{land}, though the 
summation is extended to all couples of spins and not
restricted to first neighbors. 
Following the Kac procedure \cite{hmfequi}, \index{Kac prescription} 
the coupling constant in the potential is 
divided by N. This makes $H$ only formally extensive 
($V\sim N$ when $N\rightarrow\infty$)\cite{tsallis1,tsallis2,celia,gia}, 
since the energy remains non-additive, i.e. the system cannot 
be trivially divided in two independent sub-systems.   
The canonical analytical solution of the model 
predicts a second-order phase transition. \index{phase transition}
 At low energy density  the system is in  a 
ferromagnetic phase characterized by a  magnetization  $M\sim1$, where 
  M is the modulus of  
\begin{equation}  
{\bf M} ={\frac{1}{N}}\sum_{i=1}^N {\bf m}_i ~~,
\end{equation}
with ${\bf m}_i=[cos(\theta_i), sin(\theta_i)]$. Increasing the energy 
beyond a critical value, the spins become  homogeneously oriented 
on   the unit circle and $M\sim0$. 
The {\em caloric curve}, i.e. the  
dependence of  the energy density $U = E/N$ on the temperature $T$,  
\index{caloric curve}
is  given by $U = {T \over 2} +  {1 \over 2} ( 1 - M^2 )$ \cite{hmfequi} 
and shown in Fig.\ref{calor} as a full curve. 
The critical point is  at energy density $U_c=0.75$ 
corresponding to a critical temperature $T_c=0.5$ \cite{ant1}. 
The dynamical behavior of HMF can be investigated in the microcanonical 
ensemble by starting the system with water bag initial conditions, 
i.e. $\theta_i=0$  for all $i$ ($M=1$) and velocities uniformly 
distributed, and integrating numerically the equations of motion \cite{lat1}. 
As shown in Fig.\ref{calor}, microcanonical simulations are in general in good
agreement with the canonical ensemble, except for a region below $U_c$, 
where  it has also been  found a
dynamics characterized by  L\'evy walks, anomalous diffusion \cite{lat2}  
and a negative specific heat\cite{lat3}.
Ensemble inequivalence and negative specific heat 
\index{negative specific heat} 
have also been found in self-gravitating systems \cite{lyn}, 
nuclei \index{nuclear physics} 
and atomic clusters  \cite{gro,dago,chomaz,ato}. In our case 
such anomalies emerge as dynamical transient  
features, although in a generalized version of the HMF model \cite{tor,hmfequi}
it seems that they remain also in the BG equilibrium regime. 
In order to understand better  this disagreement we focus on a particular 
energy value, namely $ U=0.69$ below the critical energy, 
and we study the time evolution 
of temperature, magnetization, and velocity distributions.
              
In Fig.\ref{voliqss} we report the temporal evolution of $2<K>/N$
for $U=0.69$ and $N=500$.
This  quantity, at equilibrium, 
is expected to coincide with the temperature 
($<\cdot>$ denotes time averages). The figure shows that when the system 
is started with out-of-equilibrium initial conditions (water bag in our case) 
rapidly reaches a metastable  quasi-stationary 
state (QSS) which does not coincide with   the canonical prediction. 
\index{non-equilibrium phenomena}
After a short transient time,not reported  here,  $2<K>/N$ shows 
a plateau corresponding to a N-dependent temperature 
$T_{QSS}(N)$ (and $M_{QSS}\sim 0$, $M\rightarrow 0$ if 
$N \rightarrow \infty$) lower than the canonical temperature.  
This metastable state needs a long time to relax to the 
canonical equilibrium state with temperature 
$T_{can}=0.476$ and magnetization $M_{can}=0.307$. 
In the same figure we plot  
a typical single-particle motion in the two insets. In panel (b) the 
motion is characterized by anomalous diffusion and L\'evy walks  with
constant velocity\cite{lat2}.  \index{anomalous diffusion}
This motion is well described by a  model introduced in 
ref. \cite{klafter}. 
These features change at equilibration, see panel (c)
where the motion is a standard random walk and diffusion becomes normal.
In ref. \cite{lat2}  it has been found that 
the lifetime of the metastable QSS and the crossover time from anomalous 
to normal diffusion do coincide within the numerical accuracy adopted. 
On the other hand, 
the duration of the plateau increases with the size of the system. In
fact the lifetime of QSS has 
a linear dependence on N as shown in  Fig.\ref{scaling} where simulations
for different out-of-equilibrium initial conditions are plotted. 
These numerical results indicate that 
the two limits  $ t\rightarrow \infty$ and $ N\rightarrow \infty$ 
do not commute: if  the thermodynamic limit \index{thermodynamic limit} 
is performed before the infinite time limit, 
the system does not relax to the BG equilibrium and the anomalies 
discussed above remain forever \index{relaxation}. Indeed similar features 
have also been found for spin glasses\cite{par,stil} 
and granular matter \cite{kud,coniglio} and could probably be 
interpreted within a general theoretical scenario.  
When N increases $T_{QSS}(N)$ tends to a definite value which is 
$T_{\infty} =0.38$. The latter  is obtained analytically 
as the metastable prolongation,
at energies just below $U_c=0.75$, of the
 high-energy branch charaterized by $M=0$. 
In ref.\cite{tsahmf1} 
we have also found that $ [T_{QSS}(N)-T_{\infty}] \propto N^{-1/3}$. This 
scaling, exploiting the expression for the caloric curves, implies also a 
scaling of the magnetization, which goes to zero like
  $M_{QSS} \propto  N^{-1/6}$.
A further numerical check of this law was obtained 
in ref.\cite{tsahmf2}. 

It is important at this point to discuss 
the behavior of the Lyapunov exponents as a function of N and 
of time in order to investigate if also this measure
shows some anomaly. In fact this is just the case: also the largest Lyapunov
exponent $\lambda$
has two temporal regimes and in particular two plateaux. We plot in   
Fig.\ref{lle} the time evolution of $\lambda$
 for the case $U=0.69$ and $N=500$.
The numerical simulation averaged over 350 events shows that in the metastable
QSS regime the largest Lyapunov exponent 
 is smaller than at the BG equilibrium. In the inset we report the 
scaling of $\lambda$ in the QSS regime as a function of the size:
increasing $N$,  $\lambda_{QSS}$ tends to zero 
as $N^{-1/9}$ \cite{tsahmf2}. The scaling law for the Lyapunov exponent 
 can be related to the scaling of the magnetization
as discussed  in ref.\cite{tsahmf2}. The fact that 
$\lambda \rightarrow 0$ for $N \rightarrow \infty$ implies 
that mixing is negligible as N increases and one expects anomalies 
in the relaxation process in the sense  advanced by Krylov \cite{krylov} 
\index{chaotic mixing} \index{relaxation}
and in a very similar way to the case of the maps discussed in the previous
section.
\begin{figure}[th]
\begin{center}
\includegraphics[width=.75\textwidth]{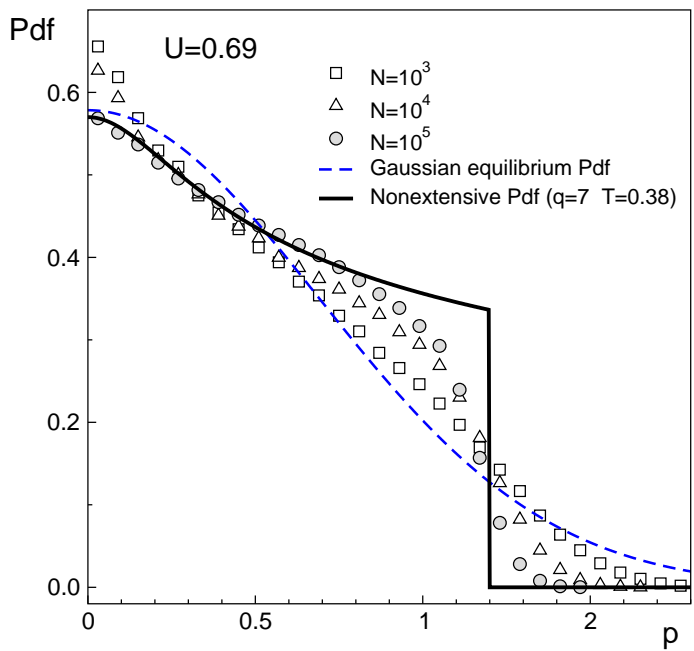}
\end{center}
\caption[]{We plot the velocity probability 
distribution functions of the HMF model for U=0.69 and different N values 
(open symbols). The Gaussian equilibrium shape is also plotted (dashed curve)
together with the fit with the nonextensive pdf formula 
obtained for $q=7$ and $T=0.38$ (full curve).   }
\label{pdf}
\end{figure}
\begin{figure}[th]
\begin{center}
\includegraphics[width=.75\textwidth]{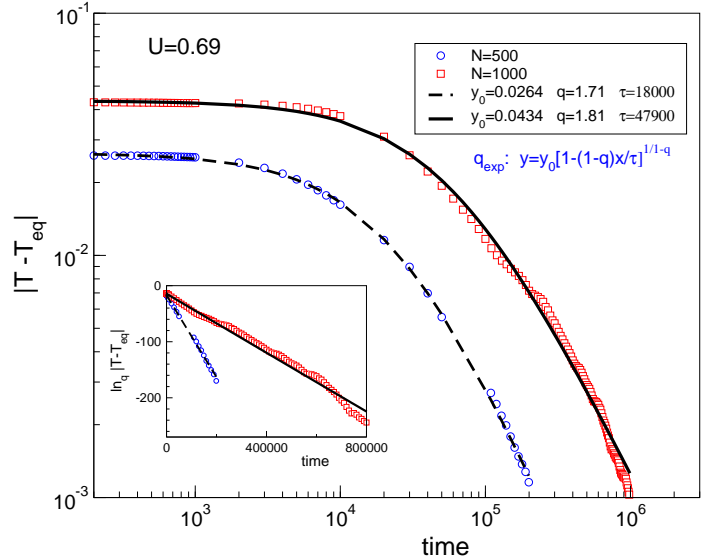}
\end{center}
\caption[]{For th HMF model, we plot the difference 
between the temperature T and that one at equilibrium $T_{eq}$,  
for U=0.69 and  N=500,1000, vs time (open symbols). The points are
the result of an ensemble average over 1000  runs.
The figure shows that the relaxation process is a power law in time 
and can be very well reproduced with a decreasing
 $q$-exponential. We  report the $q$-exponential fits 
for comparison (full and dashed curves). 
In the inset we plot the q-logarithm of the same data.
The entropic index $q$ is related to relaxation and should not necessarily
be equal to that one of velocity pdfs. }
\label{relax}
\end{figure}

In order to check relaxation \index{relaxation} 
in HMF not only the temperature, 
the second moment, but also the higher moments 
of the velocity distribution should coincide with those of
the BG equilibrium one.
In Fig.\ref{pdf} we focus our attention on the velocity 
probability distribution functions (pdfs). The figure shows that 
the initial uniform velocity pdfs, quickly acquire and mantain 
for the  complete  duration of the metastable state 
a  {\em non-Gaussian  shape }. In particular   
the velocity pdf of the QSS is wider than a Gaussian 
(drawn as dashed curve) for small velocities, but shows a faster 
decrease for $p>1.2$.
The enhancement for velocities around $ p\sim 1$  is   
consistent with the anomalous diffusion and 
the L\'evy walks (with average velocity $ p\sim 1$) 
 observed in the QSS regime \cite{lat2}. 
The following rapid decrease for $p>1.2$ is due to conservation of 
total energy. 
The stability of the QSS velocity pdf can be explained by the fact 
that, since for $N\rightarrow \infty$, 
  $M_{QSS} \rightarrow 0$  also   the force on  
the spins tends to zero with N, being 
$F_i = -M_x sin\theta_i + M_y cos\theta_i$. 
If  N is finite, one has only  
 a small random force, which makes the system  eventually 
evolve into the usual Maxwell-Boltzmann
 distribution after some time. When this happens, 
  L\'evy walks disappear and anomalous 
diffusion leaves place to  Brownian diffusion\cite{lat2}, see insets 
in Fig.\ref{voliqss}. Nonextensive thermodynamics 
is able to reproduce the non-Gaussian pdf reported  
in Fig.\ref{pdf}. In fact 
the nonextensive formalism provides, for the canonical ensemble, 
a q-dependent power-law distribution \index{power-law distribution} 
in the variables $p_i$ , 
$\theta_i$. 
This distribution has to be integrated over all $\theta_i$ and all but
one $p_i$ in order to obtain the one-momentum pdf,
$P_q(p)$, to be compared with the  numerical  one, $P_{num}(p)$, 
obtained by considering, 
within the present molecular dynamical frame, increasingly 
large N-sized subsystems 
of an increasingly large M-system. Within the $M>> N >>1$
 numerical limit, we expect to go from 
the microcanonical ensemble to the canonical one 
(the cut-off is then expected to gradually disappear as indeed 
occurs in the usual short-range Hamiltonians), thus 
justifying the comparison between $P_q(p)$ and $P_{num}(p)$. 
The enormous complexity of this procedure
 made us to turn instead onto a naive, but tractable, comparison,
 namely that of our present
 numerical results with the following one-free-particle 
pdf \cite{tsallis1,tsallis2}    
 $ P(p) \propto [1 - ({1 \over 2T}) (1-q) p^2 ]^{ 1/(1-q)}$  , which
 recovers the Maxwell-Boltzmann distribution for $q=1$. 
The same type of formula has been recently used to describe successfully 
turbulent Couette-Taylor flow \cite{bect} and non-Gaussian pdfs 
related to anomalous diffusion of {\em Hydra} cells in
cellular aggregates. 
In the case of HMF the best fit is obtained by a curve 
with entropic index $q=7$, and temperature $T=0.38$. 
The agreement between numerical results and theoretical curve improves 
with the size of the system. A finite-size scaling confirming  
the validity of the fit was reported in ref. \cite{tsahmf1}.
Since in our case we get $q>3$,  the theoretical curve does 
not have a finite integral and thus  it needs to be truncated 
with a sharp cut-off in order to make the total probability equal to one.
It is however clear that, the fitting 
value $q=7$ is only an effective nonextensive entropic index. 
In ref.\cite{tsahmf1} we verified, through the calculation of the fractal  
dimension $D_2$,  that  a 
dynamical correlation emerges in the $\mu$-space  before the final
 arrival to a quasi-uniform 
distribution. During intermediate times some filamentary structures 
appear, a similar feature
 has recently been found also in self-gravitating systems\cite{kon}, 
which might be closely related to
the plateaux observed in Fig.\ref{voliqss}. 
We have found that these correlations do not 
sensibly depend on N, thus likely 
the possible connection does  not concern the entire 
$\mu$-space, but perhaps
only the small  sticky regions between the "chaotic sea" and 
the quasi-orbits\cite{kla}. 
Recently another application of the generalized 
nonextensive formalism has been observed for this model
in the relaxation regime. It has been found that the time relaxation of
macroscopic quantities, like the temperature for example,  
obey power-laws and more specifically $q$-exponential curves.  
To illustrate this, in Fig.\ref{relax}, \index{relaxation}
 we plot the time relaxation of the temperature to its final 
value for $U=0.69$ and for two sizes N=500,1000. 
In particular we show $y(t)= | T - T_{eq}  |$, with $T_{eq}=0.476 $ as 
a function of the time $t$. The system relaxes to equilibrium 
according to the $q$-exponential formula\footnote {The minus 
in front of the term $(1-q)$ at variance with  formula (\ref{qexp}) 
is due to the fact we are considering a decreasing $q$-exponential and $q>1$.}
 \begin{equation}  
  y(t) = y_0 \left[1 - \left(1 -q \right) 
{t\over \tau } \right]^{ 1/(1-q)}  ~~,
\end{equation} 
where $q$ is the entropic index, 
$\tau$ a characteristic time and $y_0$ a saturation
value \cite{tsallis1}. 
For $N=500$ the same data shown in Fig.\ref{voliqss} are used. 
For this particular  size, one finds  a value of the entropic index
equal to $q=1.71$ and a characteristic relaxation time $\tau=18000$.
A slightly greater value of $q$ equal to $1.81$ 
is found for N=1000.
In the inset we report the q-logarithm of same data \cite{rapis}. 
This entropic index $q$ is related to the relaxation and there is no reason
to be exactly the same as the one obtained for the velocity pdfs.
It is interesting to notice that similar relaxation process 
characterized by a $q$-exponential has been recently discovered 
also in echo experiment on the Internet \cite{abe}.  
Two final considerations before concluding this section. The first one
is that in ref.\cite{tama} it has been found that the HMF model
shows aging in correspondence of the metastable QSS. 
Thus it is a very intriguing challenge left for future investigation, 
the study of the possible connections between 
spin glasses, hamiltonian mean field models and nonextensive statistics. 
Second, the anomalies we have found are not a peculiar characteristic of
the HMF model. In fact the same behavior has been found also
in a generalization of the HMF model, the $\alpha-XY$
model \cite{celia,gia}, in which the spins are placed on a
d-dimensional lattice, and the interaction energy between
two generic spins $i$ and $j$ (of formula (\ref{hmfpot}))
is modulated by an extra factor $1/r_{ij}^{\alpha}$.
Here $r_{ij}$ is the distance between spin $i$ and spin $j$
 \cite{gia}. If $\alpha/d <1 $ nonextensivity induces
anomalies similar to those described for the HMF model. 
Conversely, if
$\alpha/d>1$, i.e. when the potential decreases rapidly enough with
$r$, the $\alpha-XY$ model behaves in the standard way
 and does not exibit any of
the anomalies discussed in this section. 
Similar features  induced by nonextensivity 
have been found also for long-range Lennard-Jones-like  
potentials \cite{borges}.

\section{ Conclusions}
We have finally ended our guided tour in the realm of nonextensive
systems. Our choice was to focus only on a few simple examples,
chosen in order to represent the most instructive  classes of systems.
On one hand we have considered {\it low-dimensional systems},
in particular low-dimensional maps as the logistic map
(representative of a dissipative system), and the
standard map (representative of conservative systems).
On the other hand we have  considered {\it high-dimensional systems}
discussing the anomalies observed in the HMF model, a
classical many-body long-range-interacting Hamiltonian.
Though representing very different kind  of systems, 
all the examples reported share some common properties: long-range
correlations in space/time, either long-range microscopic interactions or
long-range microscopic memory (nonmarkovian processes), or (multi)fractal
boundary conditions, or, generically speaking, some mechanism which
creates a scale-invariant hierarchical structure of some sort.
Such properties appear in classical many-body Hamiltonians
and low-dimensional nonlinear dynamical systems
everytime that the Lyapunov exponents approach zero,
i.e. when  chaos (strong mixing) is
impossible and the sensitivity to the initial conditions
is not exponential in time.
In most cases of weak sensitivity to initial conditions,  
a generalized nonextensive formalism may replace
standard thermostatistics.

\end{document}